\documentclass{elsart}
\usepackage{natbib}
\usepackage{epsfig}
\usepackage{graphicx}
\usepackage{amssymb}

\begin{document}
\runauthor{Xu, Zhou}
\begin{frontmatter}

\title{Quarks and Quakes}

\author[Beijing]{R. X. Xu}\footnote{
Corresponding author.\\
{\em Email address:} rxxu@bac.pku.edu.cn (R. X. Xu).},
\author[CAS-Shanghai]{A. Z. Zhou}

\address[Beijing]{School of Physics, Peking University,
Beijing 100871, China}

\address[CAS-Shanghai]{Shanghai Astronomical Observatory,
Chinese Academy of Sciences, Shanghai 200030, China}

\begin{abstract}
A quake model of bare strange stars for normal pulsar glitches is
summarized. Three mechanisms being responsible for developing
elastic stress energy are presented. It is suggested that other
kinds of glitches (e.g, the frequency glitch in KS 1947+300 and in
AXP/SGRs) could represent the bulk-strain-induced quakes. The
apparent field increase of normal pulsars into ``magnetars'' could
be the result of stellar catastrophic shrinking when the elastic
force raises to a critical point.
\end{abstract}


\end{frontmatter}

%________________________________________________________________

%\section{Introduction}
{\bf 1. Introduction}

Is ``{\em pulsars $=$ normal neutron stars}'' the truth? Most of
astrophysicists may believe this equation, but they can {\em not}
provide sufficient evidence for the belief since {\em no} one can
rule out another equation of ``{\em pulsars $=$ quark stars with
strangeness}'' with certainty.
The concept, that the nature of pulsars is actually strange star,
is not new, and the difficulty of which is also very old
\citep{alpar87} since fluid strange-star (even with possible
crusts) models were noted to be inconsistent with the observations
of pulsar glitches more than one decade ago.
Modifications with the inclusion of possible stable particles to
form a differentiated structure of so-called strange pulsars was
also suggested \citep{benv90}, but is not popular because of a
disbelief in the employed physics \citep{hps93,horv04}.
And then, should the second equation be wrong according simply to
glitches observed?

A key point to reproduce glitches of strange stars could  be the
conjecture that strange quark matter is in a solid state when its
temperature is not as high as a few MeV \citep{xu03a,xu04}. The
reason that we try to retrieve strange stars from the disadvantage
above is that observational hints of (bare) strange stars do exist
\citep{xu03b,xu03c}.
Additionally, a solid strange star model can also explain the
discrepancy between free precession \cite{sls00,slu01} and
glitches of radio pulsars, since the the precession would be
damped out on timescales being much smaller than observed in the
neutron star model where glitches can be understood.
Link \cite{link03} points also that a neutron star core containing
coexisting neutron vertices and proton flux tubes is inconsistent
with observations of freely precessing pulsars.

{\em Can a solid strange quark star be possible?}
Certainly it is not necessary for quark matter to have
strangeness, but it is conjectured that the presence of
strangeness might be energetically favorable. Phenomenological
calculation shows that the conjecture is very likely to be true
\citep{fj84}.
For quark matter in low temperature, $T$, but high baryon density,
$n_{\rm b}$, there exists a {\em competition} between color
superconductivity and solidification, just like the case of
laboratory low-temperature physics.
Normal matter should be solidified as long as the interaction
energy between neighboring ions is much higher than that of the
kinetic thermal energy. In this sense, high density (then a small
separation between ions, $l$) and low temperature favor a solid
state. However, this classical view is not applicable when quantum
effects dominate, i.e., the de Broglie wavelength of ions
$\lambda\sim h/\sqrt{3mkT}>l$ ($m$ the mass of ions).
So small values of $l$ and $T$ can also result in a quantum state
of the matter. This is the competition!
We need thus {\em weak interaction} between particles and {\em low
mass} in order to obtain a quantum fluid before solidification.
This is why {\em only helium}, of all the elements, shows
superfluid phenomenon though other noble elements have similar
weak strength of interaction due to filled crusts of electrons.
The color interaction ({\em and} the Coulomb interaction in system
with strangeness) may be responsible for forming quark-clusters
(being similar to the $\alpha$-clusters in nucleon-freedom) in
quark matter with low $T$ but high $n_{\rm b}$.
The residual strong interaction between the quark-clusters could
conduce towards a solid state of quark stars in much low
temperature.

{\em Can a solid neutron star be possible?}
The nuclear liquid model is experimentally successful, the
original version of which is the nuclear Fermi-gas model. Nucleons
can be regarded as Fermi-gas due to the Pauli exclusive in the
model. A great part of neutron star matter has density to be
approximately the nuclear saturation density, and at least this
part is in a fluid state. The answer to the question could thus be
``no''.
Therefore, a quark star is identified if one convinces that a
pulsar is in a solid state.

One negative issue is that giant glitches are generally not able
to occur at an observed rate in a solid neutron star \citep{bp71}.
Nonetheless, more strain energy could be stored in a solid quark
star due to an almost uniform distribution of density (the density
near the surface of a bare strange star is $\sim 4\times 10^{14}$
g/cm$^3$) and high shear modulus introduced phenomenologically for
solid quark matter with strangeness. More energy is then released,
and this may enable a solid pulsar to glitch frequently with large
amplitudes.
Furthermore, the post-glitch behaviors may represent damped
vibrations.
Zhou et al. \citep{z04} has modelled glitches in a starquake
scenario of solid quark stars.

%\section{Glitches as starquakes}
{\bf 2. Glitches as starquakes}

A quake model for a star to be mostly solid was generally
discussed by Baym \& Pines \citep{bp71}, who parameterized the
dynamics for solid crusts, and possible solid cores, of neutron
stars.
Strain energy develops when a solid star spins down until a quake
occurs when stellar stresses reach a critical value.
We suggest that, during a quake, the entire stress is almost
relieved at first when the quake cracks the star in pieces of
small size (the total released energy $E_{\rm t}$ may be converted
into thermal energy $E_{\rm therm}$ and kinematic energy $E_{\rm
k}$ of plastic flow, $E_{\rm t}=E_{\rm therm}+E_{\rm k})$, but the
part of $E_{\rm k}$ might be re-stored by stress due to the
anelastic flow (i.e., the kinetic energy is converted to strain
energy again).
A quark star may solidify with an initial oblateness
$\varepsilon_0$; stress (to be relative to $\varepsilon_0$)
increases as the star losses its rotation energy, until the star
reaches an oblateness $\varepsilon_{+1}$. A quake occurs then, and
the reference point of strain energy should be changed to
$\varepsilon_1$ (the oblateness of a star without shear energy) at
this moment. After a glitch, the star solidifies and becomes
elastic body again.

The density of quark stars with mass $< \sim 1.5 M_\odot$ can be
well approximated to be uniform.
As a star, with an initial value $\varepsilon_{0}$, slows down,
the expected $\varepsilon$ decreases with increasing period.
However, the rigidity of the solid star causes it to remain more
oblate than it would be had it no resistance to shear. The strain
energy in the star reads \citep{bp71}
\begin{equation}
E_{\rm strain}= B(\varepsilon  - \varepsilon_{0} )^2,
\end{equation}
and the mean stress $\sigma$ in the star is
\begin{equation}
\sigma  = \left| {\frac{1}{{V_{} }}\frac{{\partial E_{\rm strain}
}}{{\partial \varepsilon }}} \right| = \mu (\varepsilon _0  -
\varepsilon ),%
\label{sigma}
\end{equation}
where $\varepsilon$ is stellar oblateness, $V=4\pi R^3/3$ is the
volume of the star, and $\mu  = 2B/V $ is the mean shear modulus
of the star.
Here the stress developed by the decrease of oblateness (due to
the spindown) is included at first (the stress developed by other
factors is discussed in \S3).

The total energy of a star with mass $M$ and radius $R$ is mostly
the gravitational energy $E_{\rm gravi}$, the rotation energy $ E_
{\rm rot}$, and the strain energy $E_{\rm strain}$,
\begin{equation}
E = E_{\rm gravi} + E_ {\rm rot} + E_{\rm strain}= E_{\rm 0} +
A\varepsilon ^2 + L^2/(2I) + B(\varepsilon  - \varepsilon_i)^2
\label{E}
\end{equation}
where $\varepsilon_i$ is the reference oblateness before the
$(i+1)$-th glitch occurs, $E_{\rm 0}=-3M^2G/(5R)$, $I$ is the
moment of inertia, $L=I\Omega$ is the stellar angular momentum,
$\Omega=2\pi/P$ ($P$ the rotation period), and the coefficients
$A$ and $B$ measure the gravitational and strain energies
\citep{bp71}, respectively,
\begin{equation}
A = \frac{3}{{25}}\frac{{GM_{}^2 }}{R},~~~~~
B = {2\over 3}\pi R^3 \mu.
\end{equation}
By minimizing $E$, a real state satisfies (note that $\partial
I(\varepsilon)/\partial \varepsilon=I_0$),
\begin{equation}
 \varepsilon  = \frac{{I_0 \Omega ^2}}{{4(A +
B)}} + \frac{B}{{A + B}}\varepsilon _i. %
\label{epsilon}
\end{equation}
The reference oblateness is assumed, by setting $B=0$ in
Eq.(\ref{epsilon}), to be
\begin{equation}
\varepsilon _i  = I_0 \Omega^2/(4A).%
\label{epsiloni}
\end{equation}
A star with oblateness of Eq.(\ref{epsiloni}) is actually a
Maclaurin sphere. When the star spins down to $\Omega$, the stress
develops to
\begin{equation}
\sigma = \mu [\frac{A}{{A + B}}\varepsilon_i - \frac{{I_0 \Omega^2
}}{{4(A + B)}} ],%
\label{sigmamu}
\end{equation}
according to Eq.(\ref{sigma}). A glitch occurs when
$\sigma>\sigma_{\rm c}$ ($\sigma_{\rm c}$: the critical stress).

A detail model in this scenario is introduced in Zhou el al.
\citep{z04}, where it is found that the general glitch natures
(i.e., the glitch amplitudes and the time intervals) could be
reproduced if solid quark matter, with high baryon density but low
temperature, has properties of shear modulus $\mu=10^{30\sim 34}$
erg/cm$^3$ and critical stress $\sigma_{\rm c}=10^{18\sim 24}$
erg/cm$^3$.

%\section{Why quake?}
{\bf 3. Why quake?}

The key point that a solid star differs from a fluid one is stress
energy developed only available in the solid star. Factually, an
elastic body can have two kind of changes: {\em shearing} and {\em
bulk} strains. The volume, $V$, changes in the later, but not in
the former. If both strains are included, Eq.(\ref{E}) would
become,
\begin{equation}
E=E_0+A \varepsilon^2+L^2/(2I)+B(\varepsilon  -
\varepsilon_i)^2+K(\Delta V/V_0)^2,
\label{EE}
\end{equation}
where $K$, which is order of $\mu$, is the bulk modulus, $\Delta
V=V-V_0$, and $V_0$ the volume of the body without stress.
Besides the shear strain of ellipsoid change discussed in \S2,
azimuthal stress due to the general relativistic effect (being
similar to the frame-dragging effect in vacuum) of rotating solid
stars may also contribute significantly, though, unfortunately,
the theoretical answers to elastic relativistic-stars with
rotation are very difficult to be worked out.

Elastic energy develops as a solid bare strange star cools
($\Rightarrow$ bulk strain) and spins down ($\Rightarrow$ shearing
strain; even spinning constantly).
The temperature-dependent quantity $\Delta V/V_0\sim (\Delta
R/R_0)^3$, with $R$ the radius, $V_0\simeq 4\pi R_0^3/3$, $T$ the
temperature. The value of $R(T=40{\rm MeV})-R(T=0)$ could be order
of a few hundreds of meters \citep{blaschke04}. The giant
frequency glitch in KS 1947+300 could be evidence for a quake
caused by bulk-stress energy release (i.e., bulk elastic force
increases to a critical point), but one may expect a sudden
decrease in the pulse frequency when a star is spinning up in the
glitch model of normal neutron stars \citep{galloway04}.
The glitch in KS 1947+300 can be reproduced as long as the star
shrinks with $\Delta R/R=-0.5 \Delta\nu/\nu\sim 10^{-5}$.

{\em Anomalous X-ray pulsars (AXPs) \& Soft $\gamma$-ray repeaters
(SGRs)}.
AXP/SGRs are supposed to be magnetars, a kind of neutron stars
with surface fields of order of $10^{13\sim 14}$ G, or even
higher. But an alternative suggestion is that they are
normal-field pulsar-like stars which are in an accretion propeller
phase. The very difficulty in the later view point is to reproduce
the irregular bursts, even with peak luminosity $\sim 10^7L_{\rm
Edd}$ (SGR 0526-66; $L_{\rm Edd}$ the Eddington luminosity).
Though it is possible that giant bursts may be the results of the
bombardments of comet-like objects (e.g., strange planets) to bare
strange stars, moderate bursts could be of quake-induced.
Both shear and bulk strain-induced quakes could occur in AXP/SGRs
when the stress energy increase to a critical value. Stress energy
as well as magnetic energy (and probably gravitational energy)
could be released during quakes.
The glitches in AXP/SGRs (e.g., 1E 2259+586, around the 2002
outburst) could be examples of such quakes \citep{kaspi04}.
For $\sigma_c\sim 10^{22}$ erg/cm$^3$, the total elastic energy
released could be order of $\sim 10^{40}$ erg/cm$^3$ when a quake
occurs in a solid quark star with radius $\sim 10^6$ cm. This
energy is comparable to the observed values for two X-ray flares
in 1E 1048.1-5937 \citep{gk04}.

It is suggested \citep{lyne04,lz04} that pulsar-like stars could
evolve from normal radio pulsars to magnetars (AXP/SGRs) through
un-recoverable glitches (neither the period nor the period
derivative are completely recovered for Crab, Vela, and other
young pulsars).
However, the ``appearance'' field increasing could arise from the
shrinking of pulsars after quakes. If the effect of decreasing $I$
is included, the pulsar field, $B$, should be derived by solving
\begin{equation}
I{\dot \Omega} + {1\over 2}{\dot I}\Omega^3= -{2\over 3c^3}B^2R^6\Omega^3.%
\label{dotO}
\end{equation}
%

%\section{Conclusions \& Observational tests}
{\bf 4. Conclusions \& Observational tests}

We have summarized the quake model for pulsar glitches. The
elastic stress energy develops by spindown, cooling, and the
frame-dragging effect. The bulk-strain-induced quakes may cause
glitches in KS 1947+300 and in AXP/SGRs, as well as the apparent
field increase of normal pulsars into ``magnetars''.

Gravitational wave detection could be an useful tool to test the
quake model \citep{xu03a}. In addition, candidates for strangelets
known in the literature could also teach us that one should update
neutron stars by strange stars since cosmic strangelets would
render neutron stars unstable to being strange stars.

{\em Acknowledgments}:
This work is supported by National Nature Sciences Foundation of
China (10273001), and by the Special Funds for Major State Basic
Research Projects of China (G2000077602).

\end{document}